\documentclass[3p]{elsarticle}
\pdfoutput=1
\usepackage{graphicx}
\usepackage{caption}
\usepackage{subcaption}
\usepackage{url}
\usepackage{amsmath,amssymb}
\usepackage[T1]{fontenc}
\usepackage{ae,aecompl}
\usepackage{xcolor}
\usepackage{textcomp}
\usepackage[english]{babel}
\usepackage{listings}

\begin{document}

\begin{frontmatter}

\title{Distributed Multiscale Computing with MUSCLE~2, the Multiscale Coupling Library and Environment}
\author[uva]{J. Borgdorff\corref{joris}}
\ead{J.Borgdorff@uva.nl}
\author[psnc]{M. Mamonski}
\ead{mamonski@man.poznan.pl}
\author[psnc]{B. Bosak}
\ead{bbosak@man.poznan.pl}
\author[psnc]{K. Kurowski}
\ead{krzysztof.kurowski@man.poznan.pl}
\author[unige]{M. Ben Belgacem}
\ead{mohamed.benbelgacem@unige.ch}
\author[unige]{B. Chopard}
\ead{bastien.chopard@unige.ch}
\author[ucl]{D. Groen}
\ead{d.groen@ucl.ac.uk}
\author[ucl]{P.V. Coveney}
\ead{p.v.coveney@ucl.ac.uk}
\author[uva]{A.G. Hoekstra}
\ead{A.G.Hoekstra@uva.nl}

\cortext[joris]{Corresponding author}

\address[uva]{Computational Science, Faculty of Science, University of Amsterdam, Amsterdam, the Netherlands}
\address[psnc]{Pozna\'n Supercomputing and Networking Center, Pozna\'n, Poland}
\address[unige]{Computer Science Centre, University of Geneva, Carouge, Switzerland}
\address[ucl]{Centre for Computational Science, University College London, London, United Kingdom}

\begin{abstract}

We present the Multiscale Coupling Library and Environment: MUSCLE~2. This
multiscale component-based execution environment has a simple to use Java, C++,
C, Python and Fortran API, compatible with MPI, OpenMP and threading codes. We
demonstrate its local and distributed computing capabilities and compare its
performance to MUSCLE~1, file copy, MPI, MPWide, and GridFTP. The local
throughput of MPI is about two times higher, so very tightly coupled code
should use MPI as a single submodel of MUSCLE~2; the distributed performance of
GridFTP is lower, especially for small messages. We test the performance of a
canal system model with MUSCLE~2, where it introduces an overhead as small as
5\% compared to MPI.

\end{abstract}

\begin{keyword}
distributed multiscale computing\sep multiscale modelling\sep model coupling\sep execution environment\sep MUSCLE
\end{keyword}

\end{frontmatter}

\section{Introduction}

Multiscale modelling and simulation is of growing interest~\cite{Groen:2012vf},
with appeal to scientists in many fields
such as computational biomedicine~\cite{Sloot:2010ih}, biology
\cite{Southern:2008vm}, systems biology~\cite{Dada:2011vc},
physics~\cite{E:2007ve}, chemistry~\cite{Ingram:2004wj} and earth
sciences~\cite{Armstrong:2009ft}. Meanwhile, there are efforts to provide a more
formal way of describing multiscale models~\cite{Ingram:2004ba,Yang:2009vm,Borgdorff:2012uq},
including our Multiscale Modeling and Simulation Framework
(MMSF)~\cite{Chopard:2011jn,Hoekstra:2010uv,Hoekstra:2007er,Borgdorff:2011tl}.

This framework describes the process of constructing a multiscale model by
identifying and separating its scales. It then provides a computational modelling
language and environment to create and deploy such models on a range of
computing infrastructures. For an example of biomedical applications in this context,
see~\cite{Groen:2013jn}. The Multiscale Modeling Language (MML), part of the
Multiscale Modeling and Simulation Framework, facilitates a formal
characterization of a multiscale model, and specifies how
it can be computed~\cite{Falcone:2010to,Borgdorff:2012uq}. In this paper we
present a means to implement multiscale models: the Multiscale Coupling Library
and Environment 2 (MUSCLE~2). It takes a component-based approach to multiscale
modelling, promoting modularity in its design. It is open source software under
the LGPL version 3 license and is available at~\url{http://apps.man.poznan.pl/trac/muscle}.
MUSCLE~1~\cite{Hegewald:2008vp} generally
had the same architecture and it was based on the Complex Automata
theory~\cite{Hoekstra:2010uv,Hoekstra:2007er} and focussed on multi-agent
multiscale computing; their differences are discussed in Appendix~\ref{appendix:muscle1}.

Distributed computing is a way to take advantage of limited and heterogeneous
resources in combination with heterogeneous multiscale models. There are several
motivations for distributing the computation of a multiscale model: to make use
of more resources than available on one site; making use of heterogenous resources
such as clusters with GPGPUs, fast I/O, highly interconnected CPU's,
or fast cores; or making use of a local software
license on one machine and running a highly parallel code on a high-performance cluster.
Projects such as EGI and PRACE make distributed infrastructure available, and
software that uses it is then usually managed by a middleware layer~\cite{Zasada:2012es}. The
QCG-Coordinator~\cite{bosak2012new} middleware, for example, allows users to use
computational resources at multiple locations in Europe, and it supports
MUSCLE~2 for this purpose.

Quite a few open and generic component based computing frameworks already exist, for
instance the CCA~\cite{Allan:2004va} with CCaffeine~\cite{Allan:2005wk}, the Model
Coupling Toolkit (MCT)~\cite{Kim:2011eb,Larson:2001vx}, Pyre~\cite{Pyre}, or
OpenPALM~\cite{OpenPALM}; see the full comparison by Groen~\emph{et~al.}~\cite{Groen:2012vf}.
The Model Coupling Toolkit has a long track-record and uses Fortran code
with MPI as a communication layer so it potentially makes optimal use of
high-performance machines. OpenPALM uses TCP/IP as a communication layer and
it is packaged with a graphical user interface to couple models.
Both frameworks provide some built-in data transformations.
MUSCLE~2 uses shared memory for models started in the same command and
TCP/IP for multiple commands. An advantage over the other mentioned frameworks
is that it provides additional support for distributed computing and for Java.
However, it has fewer built-in data transformations available.

There are many libraries for local and wide-area communications, apart
from MPI implementations and the ubiquitous TCP/IP sockets.
MPWide~\cite{Groen:2010gw}, for instance, is a lightweight library that
optimizes the communication speed between different
supercomputers or clusters; ZeroMQ~\cite{ZeroMQ} is an extensive communication
library for doing easy and fast message passing. To use them for model coupling
these libraries have to be called in additional glue code. MUSCLE~2 optionally
uses MPWide for wide area communication because of its speed and few dependencies.

So far MUSCLE~2 is being used in a number of multiscale models, for instance a
collection of parallel Fortran codes of the Fusion community
\cite{Frauel:2012it}, a gene regulatory network simulation
\cite{Swain:2005bk}, a hydrology application \cite{BenBelgacem:2012iu}, and in a
multiscale model of in-stent restenosis \cite{Caiazzo:2011jl,Borgdorff:2012ge,Groen:2013jn}.

In this paper, we introduce the design of MUSCLE~2 in
Section~\ref{sec:design}, including the theoretical background of the Multiscale
Modeling and Simulation Framework, MUSCLE~2's API (Application Programming
Interface) and runtime
environment. The performance and startup overhead of MUSCLE~2 is measured in
Section~\ref{sec:performance} in a number of benchmarks. Finally, in
Section~\ref{sec:usecases} two applications that use MUSCLE
are described, principally a multiscale model of a complex canal system, for which
additional performance tests are done.

\begin{figure}
	\centering
	\includegraphics[width=1\textwidth]{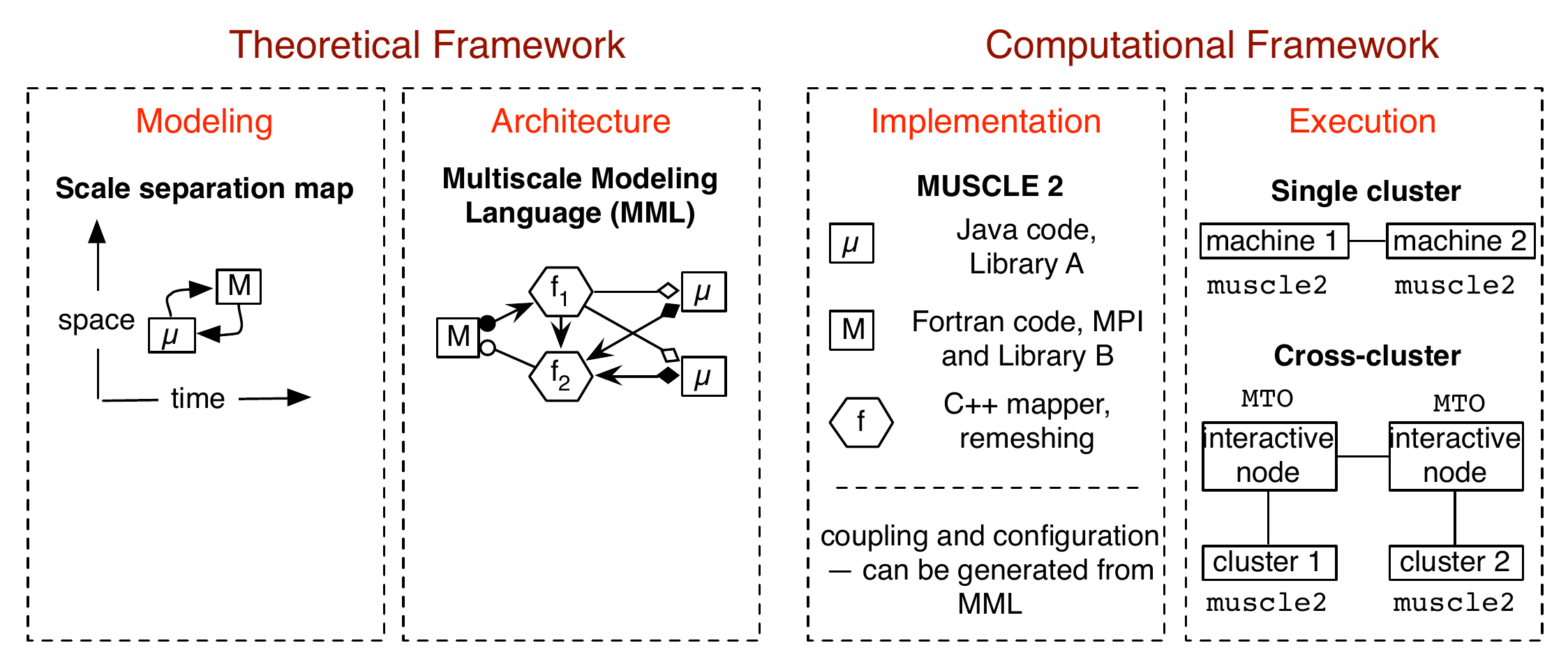}

\caption{Overview of the approach presented in this paper. First, use the
multiscale modelling and simulation framework to characterize a multiscale
model: create a scale separation map (for example with macro model M and micro
model $\mu$), and translate it to the computationally oriented multiscale
modelling language, including any mappers $f_1$ and $f_2$ to do scale bridging
and/or data conversions (Sec. 2.1). The architecture is implemented and coupled
with MUSCLE~2 (Sec. 2.2 and 2.3), and finally, executed on a single machine or
cluster with plain MUSCLE~2 (Sec 2.4), or cross-cluster using the MUSCLE
Transport Overlay (MTO, Sec. 2.5).}\label{fig:mmsf}

\end{figure}

\section{Design}\label{sec:design}
MUSCLE~2 is a platform to execute time-driven multiscale simulations. It is
component-based and thereby takes advantage of the separation between the single
scale models that together form the overall multiscale model. The components
individually keep track of the local time, and synchronize time when exchanging
messages.

A strict separation of submodels is assumed in the design of MUSCLE~2, so the
implementation of a submodel does not dictate how it should be coupled to other
submodels. Rather, each submodel sends and receives messages with specified
ports that are coupled at a later stage. At that point modellers face the
main scientific challenge: to devise and implement a suitable scale bridging
method to couple several single scale models. MUSCLE~2 supports the technical
side by offering several functional components, described in Sec.~\ref{sec:theory}.

The runtime environment of MUSCLE~2 executes a coupled multiscale model on
given computational resources. It can run each submodel on an independent
desktop machine, local cluster, supercomputer, or run all submodels at
the same location. For instance, when one or more submodels have high
computational requirements or require alternate resources such as GPU
computing, these submodels can be executed on a suitable machine, while the
others are executed on a smaller cluster. A requirement is that a connection
can be established between submodels, and that a message can only be sent to
currently running submodels. For some models a local laptop, desktop or cluster
will suffice; MUSCLE 2 also works well in these scenarios. Technical details
about the runtime environment can be found in
Appendix~\ref{appendix:technical}.

MUSCLE 2 is separated into an API, which the submodels implement, a coupling
scripting environment that specifies how the submodels will be connected, and a runtime 
environment, that actually executes the multiscale model on various resources.
The library is independent from the coupling, which is in turn independent from
the runtime environment. As a result, a submodel is implemented once and can be
coupled in a variety of ways, and then executed on any machines that are
currently available. Additionally, enhancements to the runtime environment are
possible without changing the library.

\subsection{Theoretical background}\label{sec:theory}
To generally couple multiscale models, a framework describing the foundations
of multiscale modelling~\cite{Chopard:2011jn,Borgdorff:2011tl,Ingram:2004wj,Ingram:2004ba} and its
repercussions on multiscale computing~\cite{Borgdorff:2012uq,Falcone:2010to}
was conceived. It starts by decomposing a phenomenon into multiple single scale
phenomena using a scale separation map. Based on these phenomena, single scale
models are created; see Fig.~\ref{fig:mmsf}. Ideally,
these single scale models are independent from each other, relying only on
messages at certain input ports and sending observations of their state at
output ports. By coupling the inputs and outputs of several single scale models
using so-called conduits, a multiscale model is formed. Assuming a time-driven
simulation approach, each observation and input is associated with a time
point, which should be kept consistent between single scale models. The MMSF makes
a distinction between acyclicly and cyclicly coupled models. In the former,
no feedback is possible from one submodel to the other, while in the latter a
submodel may give feedback as often as needed. This distinction has many computational
implications, such as the need to keep submodels active in cyclicly coupled models,
or the recurring and possibly dynamic need for computing resources.

To facilitate consistency, each submodel is defined with a submodel execution
loop, consisting of initialization, a loop with first an observation and then a
solving step, and then a final observation. This loop can be restarted as long
as a submodel with a coarser time scale provides input for the initial
condition. During initialization and solving steps, only input may be
requested, and during the observations, only output may be generated.

Submodels should remain independent and as such the data expected by a submodel
will not automatically match the observation of another. For this purpose data
can be modified in transit, thus implementing scale bridging methods, either by
light-weight conduit filters, which
change data in a single conduit, or by mappers, which may combine the data of
multiple sources or extract multiple observations from a single message.
Finally, the input for a submodel may not be available from another submodel
but rather from an external source, or conversely, an observation might only be
used outside the model. In that case, terminals may be used: sources to provide
data and sinks to extract data.

Each of the concepts mentioned in this paragraph is defined in the multiscale
modelling language (MML), with both a graphical representation (gMML) and an XML
representation (xMML). In MML these concepts are well-defined and accessible for
automated analysis, for example to predict deadlocks or the total runtime of a
simulation. The configuration file of MUSCLE 2 can also be generated from an
xMML file.

\subsection{Library}


The library part of MUSCLE 2 consists of Java, C, C++, Python, and Fortran APIs
and coupling definitions in Ruby. The API's for these languages can each: send
and receive arrays, strings, and raw bytes; do logging; and stage in- and
output files. Send calls have non-blocking semantics whereas receive calls are
blocking by default but may be used as non-blocking instead. The Java API, in
addition to the API's of the other languages, allows implementing formal MML
constructs such as formal submodels, filters, and mappers, and sending and
receiving advanced data structures like Java classes. Because MUSCLE~2 allows
multiple languages in a multiscale model, filters and mappers can also be used
in models that otherwise don't use Java. An example of sending and receiving
data with MUSCLE~2 is shown in Table~\ref{table:api:receive}. In all cases, the
API is non-invasive and does not force a certain programming paradigm, which
makes it straightforward to incorporate in existing code.



\begin{table}
\caption{Sending (first row) and receiving (second row) a message in MUSCLE 2 in
various programming languages.}\label{table:api:receive}\label{table:api:send}
\small
\sffamily
\begin{tabular}{l|l|l}
\rmfamily Java & \rmfamily C++ & \rmfamily Fortran\\
\hline
double[] dataA=new double[100];&
double* dataA=new double[100];&
real*8 :: dataA(1:100)\\

out(''portA'').send(dataA);&
muscle::env::send(&
call MUSCLE\_Send(\\

&
\hspace{10 pt}''portA'', dataA, 100,&
\hspace{10 pt}'portA', dataA, \footnotesize\%REF(100),\\

&
\hspace{10 pt}{\footnotesize MUSCLE\_DOUBLE});&
\hspace{10 pt}{\footnotesize\%REF(MUSCLE\_DOUBLE)})\\

&
delete [] dataA;&
\\
\hline
&
size\_t len;&
integer :: len\\

double[] dataB = (double[])&
double* dataB = (double*)&
real*8 :: dataB(1:100)\\

\hspace{10 pt}in(''portB'').receive();&
\hspace{10 pt}muscle::env::receive(''portB'',&
len = 100\\

&
\hspace{14.5 pt}{\footnotesize(void*)0},len,{\footnotesize MUSCLE\_DOUBLE});&
call MUSCLE\_Receive(\\

&
muscle::env::free\_data(&
\hspace{10 pt}'portB', dataB, len,\\

&
\hspace{10 pt}dataB, {\footnotesize MUSCLE\_DOUBLE});&
\hspace{10 pt}{\footnotesize\%REF(MUSCLE\_DOUBLE)})\\
\end{tabular}
\end{table}

\subsection{Configuration}

The configuration of a multiscale model and the coupling is done in a Ruby
file. In this file, submodels and their scales are specified, parametrized, and
coupled to each other. Mappers, conduit filters, sources and sinks are also
added to the coupling topology here. A conduit can be configured with multiple
filters; predefined-filters such as data compression, or custom filters such as
data transformations or conversions. Because the configuration is a Ruby script
the coupling topology can be automatically generated, for instance to set up a
ring or grid topology, or to read a network from a file.




Below is an example of the configuration of a multiscale model with one
macro-model and one micro-model, with a single coupling from macro to micro.

\begin{verbatim}
# Add the classpath of the submodels, using environment variable $MODEL_HOME
add_classpath ENV['MODEL_HOME'] + '/build/classes'

# Create a submodel instance 'macro' with implementing Java class 'mypackage.Macro'
macro = Instance.new('macro', 'mypackage.Macro')

# Set the macro timestep to 1 hour, and the total simulation time to 1 day
macro['dt'] = '1 hour'
macro['T'] = '1 day'

# For 'micro', use a predefined MUSCLE MPI submodel and specify the executable
micro = Instance.new('micro', 'muscle.core.standalone.MPIKernel')
micro['command'] = ENV['MODEL_HOME'] + '/build/micro'

# Couple the port 'macroscopicVariable' of macro to port 'environmentValue' of micro
macro.couple(micro, 'macroscopicVariable' => 'environmentValue')
\end{verbatim}

\subsection{Runtime environment}\label{sec:runtime}
The runtime environment of MUSCLE~2 is designed to be light-weight and portable,
and to provide high performance. MUSCLE~2 is supported on Linux and OS X. Data
is transmitted between submodels and mappers, both called instances, using
direct and thus decentralized message passing.

\begin{figure}[p]
  \centering
  \includegraphics[width=0.75\textwidth]{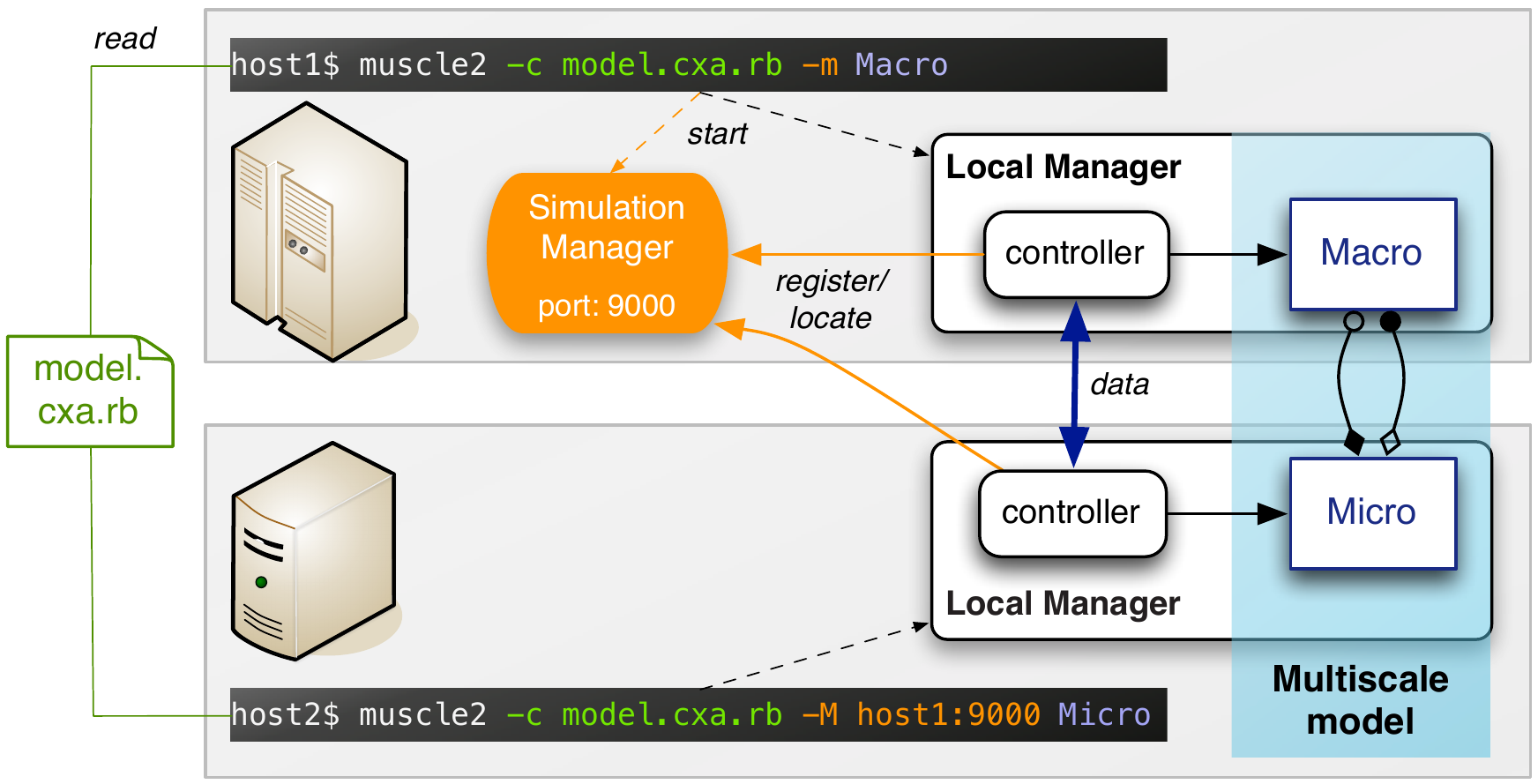}
  \caption{An example of the MUSCLE runtime environment, explained in
  Section~\ref{sec:runtime}, doing a distributed execution of the multiscale
  model described in file {\tt model.cxa.rb} on machines
  {\tt host1} and {\tt host2}. The register and data arrows are both TCP/IP
  connections. The Macro and Micro rectangles make up a multiscale model.}\label{fig:runtime:detailed}
\end{figure}

Each MUSCLE~2 simulation has a single Simulation Manager and one or more Local
Managers, as shown in Fig.~\ref{fig:runtime:detailed}. The Simulation Manager
keeps track of which instances have started and what their location is. The
Local Manager starts the instances that were assigned to it in separarate threads and
listens for incoming connections. Instances will start computation immediately but
they will block and become idle as soon as they try to receive a message that
has not yet been sent, or try to send a message to an instance that has not been
started.

A so-called native instance is a compiled instance
that runs as a separate executable. Its controller is still
implemented in Java and therefore the executable will try to establish a TCP/IP
connection with this controller, which will then do all communication
with other instances and with the Simulation Manager. A native instance may be serial or
use threading, OpenMP, or MPI.

\begin{figure}
  \centering
  \begin{subfigure}[b]{0.40\textwidth}
          \centering
          \includegraphics[width=\textwidth]{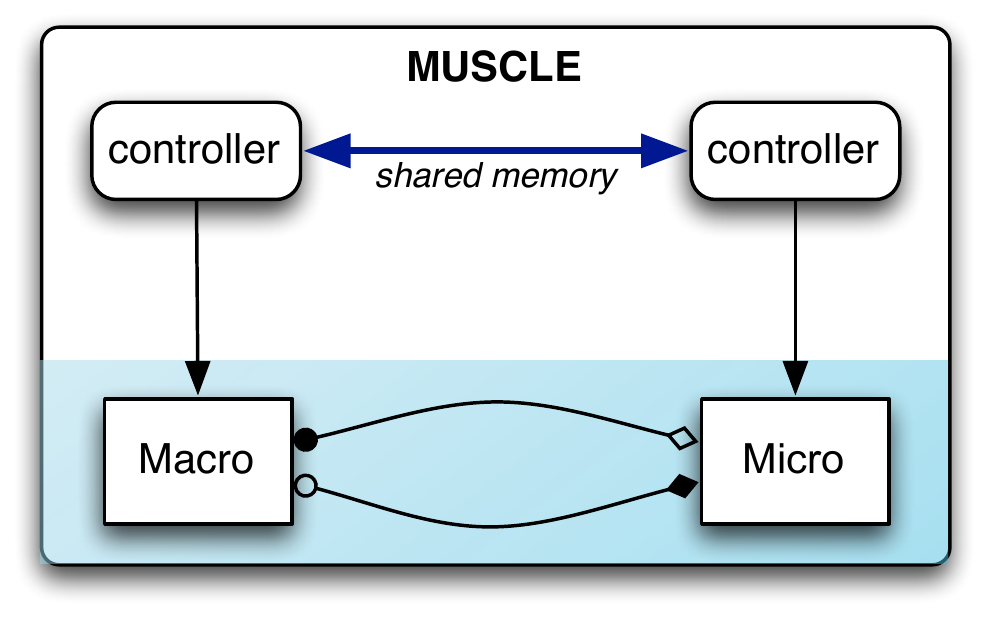}
          \caption{Single Local Manager}
          \label{fig:java:single}
  \end{subfigure}%
  ~ 
  \begin{subfigure}[b]{0.40\textwidth}
          \centering
          \includegraphics[width=\textwidth]{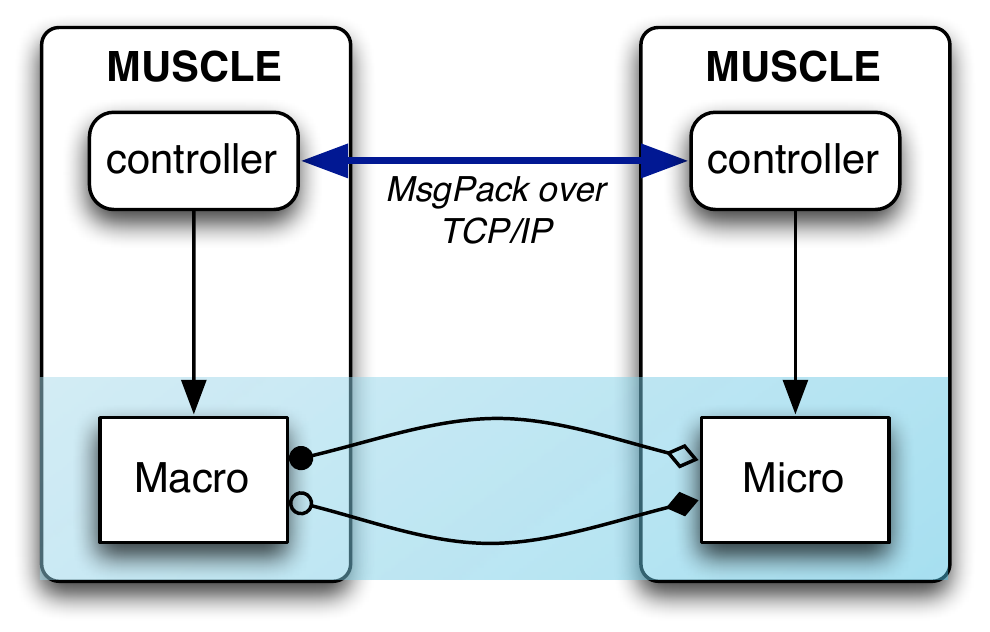}
          \caption{Two Local Managers}
          \label{fig:java:double}
  \end{subfigure}\\
  ~~~~~\begin{subfigure}[b]{0.42\textwidth}
          \centering
          \includegraphics[width=\textwidth]{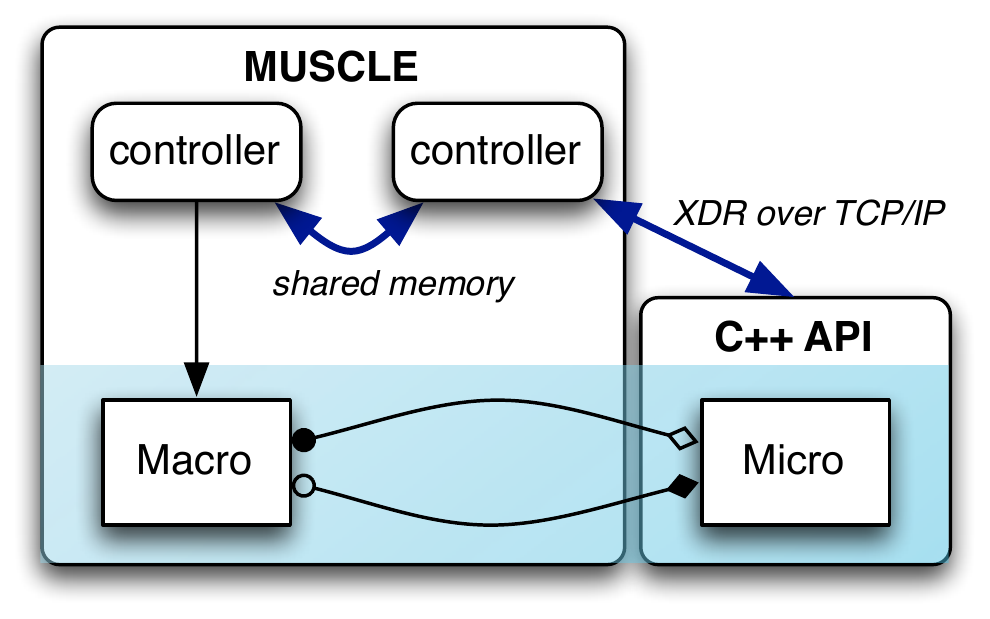}
          \caption{C++ with single Local Manager}
          \label{fig:c++:single}
  \end{subfigure}~%
  \begin{subfigure}[b]{0.43\textwidth}
          \centering
          \includegraphics[width=\textwidth]{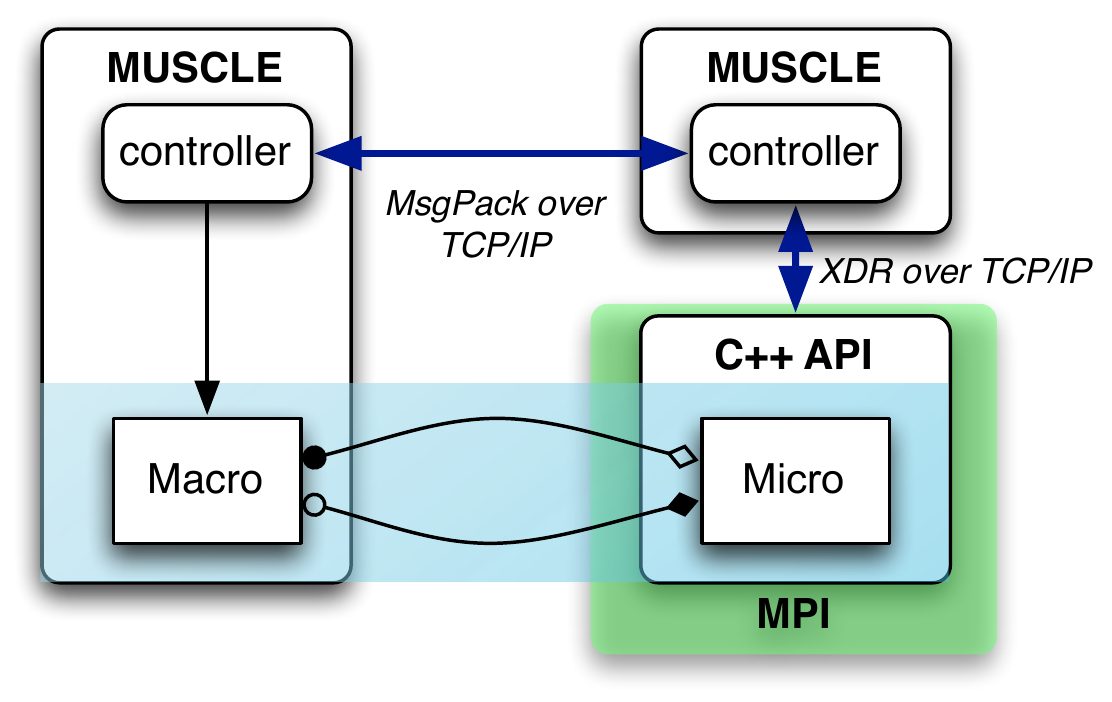}
          \caption{MUSCLE 2 with MPI}
          \label{fig:mpi}
  \end{subfigure}
\caption{Different ways in which MUSCLE 2 can be executed. A box with the MUSCLE
label indicates a Local Manager. Blue arrows indicate communication between
instances. They are labeled with the means of communication.}\label{fig:runtime}

\end{figure}

Message-passing mechanisms that are used are shown in Fig.~\ref{fig:runtime}.
Messages within a Java Virtual Machine are sent using shared memory. To insure
independence of data between instances, the data is copied once before it is
delivered from one instance to the other unless otherwise specified. Messages
between Local Managers and between the Local and Simulation Managers are sent
over TCP/IP, which is available everywhere and inherently allows messages to be
sent to other machines. The MessagePack serialization
library~\cite{MessagePack} is used for these communications because of its
efficient packing. The connection between a native instance and its controller
uses the XDR~\cite{XDR} serialization library because it is already installed
in most Unix-like systems.

\subsection{Cross-cluster computing}

Because MUSCLE~2 uses TCP/IP for message passing between instances, it can
communicate over the internet and within clusters. However, most HPC systems
prevent direct communication between submodels running on different clusters.
Therefore, MUSCLE~2 provides the user space daemon MUSCLE Transport Overlay
(MTO). It runs on an interactive node of an HPC cluster and will forward data
from MUSCLE~2 instances running on the cluster to the MTO of another cluster,
which will forward it to the intended MUSCLE~2 instance. By default, it does
this with plain non-blocking TCP/IP sockets, but it can also use the
MPWide~1.8~\cite{Groen:2010gw} library. MPWide's goal is to optimize the
performance of message-passing over wide-area networks, especially for larger
messages.

Because the MUSCLE~2 instances that make up a distributed simulation have to
run concurrently and their in- and output files have to be managed,
cross-cluster simulations are tedious to do manually. For this reason the
MUSCLE~2 framework was integrated with the QosCosGrid (QCG) middleware
stack~\cite{bosak2012new}. The QCG middleware stack offers users advanced job
management and resource management capabilities for e-Infrastructure. It will
manage the execution of a distributed MUSCLE~2 simulation from a central
location, reducing the input and management needed from the user.
\section{Performance}\label{sec:performance}

The main benefit of MUSCLE is the library and coupling environment that it
provides. However, for many if not all multiscale simulations, performance is
equally important. The performance of MUSCLE has two aspects: the overhead it
introduces and the messaging speed that it provides. These were measured on
four resources: an iMac (a local desktop machine), Zeus (a community cluster
accessible through EGI or PL-GRID), Huygens (a PRACE Tier-1 resource from
SurfSARA, the Netherlands) and SuperMUC (a PRACE Tier-0 resource from
Leibniz-Rechenzentrum, Germany). See Table~\ref{table:resources} for their details.

\begin{table}
  \caption{Computing resources used in performance testing.}\label{table:resources}
  \footnotesize
  \begin{tabular}{llllll}
    \hline
        Name & Infrastructure &                    Location &           Processor &  Cores per node\\
	\hline
        iMac &  local desktop &  Amsterdam, The Netherlands &    Intel i3 3.2 GHz &   2/4*\\
        Zeus &   PL-Grid, EGI &              Krakow, Poland &  Intel Xeon 2.4 GHz &      4\\
    SuperMUC &   PRACE Tier-0 &             Munich, Germany &  Intel Xeon 2.7 GHz & 16/32*\\
     Huygens &   PRACE Tier-1 &  Amsterdam, The Netherlands &  IBM Power6 4.7 GHz & 32/64*\\
     Cartesius &   PRACE Tier-1 &  Amsterdam, The Netherlands &  Intel Xeon 2.7 GHz & 12/24*\\
     Gordias &  local cluster &         Geneva, Switzerland &  Intel Xeon 2.4 GHz &     8\\
     Scylla &  local cluster &          Geneva, Switzerland &  Intel Xeon 2.4 GHz &     12\\
    \hline
  \end{tabular}\\
  \centering
  {\footnotesize * Two HyperThreads per core}
\end{table}

\subsection{Overhead}

With MUSCLE's runtime overhead figures, a user can estimate what the impact of MUSCLE will
be on the execution time and memory for a given multiscale model. To test the overhead
we will start a varied number of submodels and conduits, to evaluate their
impact on CPU and memory usage.
The overhead will be measured on an iMac and on SuperMUC.

To test the overhead in execution time, MUSCLE is started with 30 different submodel counts
$n$, from 1 to 1000, and 36 different conduit counts $m$, from 0 to 50.000.
The submodels are created in a configuration script, in which each
submodel adds a conduit to each of the following $m/n$ submodels, wrapping
around to the first submodel if there are less than $m/n$ succeeding submodels.
Once the simulation has started, each submodel sends and receives one empty
message through each conduit, and then exits. This way, all submodels must be
active simultaneously for a small amount of time, like they would be in a
normal simulation. Since a submodel with
native (C/C++/Fortran) code needs to start an additional executable,
it is measured separately. The amount of time spent on Java garbage collection is not
separately measured. Software versions on iMac are Java 1.7.0\_7 and Ruby 1.9.3;
on SuperMUC they are Java 1.6.0\_32 and Ruby 1.8.7. The minimal overhead $a$ is determined
by taking the minimum value encountered. The additional runtime overhead $b$ per submodel and $c$
per conduit is determined by fitting the data to the equation $a + b n + c m$,
where $n$ are the number of submodels and $m$ are the number of conduits.
The additional runtime per native submodel was fitted to a linear curve separately.
The minimum memory overhead was taken as the memory of running with one submodel,
all other values were separately fitted to a linear curve.

The results for this experiment are listed in Table~\ref{table:runtime}.
With the highest number of submodels and conduits (1000 and 50.000 respectively),
execution took 7.1 seconds on the iMac and 6.6 seconds on SuperMUC; the lowest
runtimes were 0.68 and 1.2
seconds respectively.
For most if not all multiscale simulations, even 7.1 seconds overhead will not
be significant
compared to the overhead of running the simulation, and for multiscale simulations
with less than 10 submodels, the overhead will be close to a second.

The memory consumption was measured in a similar way as runtime overhead,
except here ten submodel counts from 1 to 1000 where used, and
separately thirteen conduit counts from 0 to 50,000, each started
four times.
The Java Virtual Machine of the Local Manager was set up with an initial heap size of
1 GB and with a maximum heap size of 3 GB. Since
MUSCLE uses Java and Ruby, exact memory consumption will differ
per execution and it will include free space that their respective runtime engines
have reserved. However, with enough memory allocation a trend does emerge.
If multiple MUSCLE instances are started for a single multiscale model, additional
buffers need to be reserved for communicating with other Local Managers.
Therefore ports that are coupled to a port of a submodel with another Local Manager
are measured separately, as are submodels with native code.

The results are listed in Table~\ref{table:runtime}. With these figures, and
taking into account the memory consumption of the individual submodels, a user
can estimate how many submodels will fit in the memory of a single machine.
As a result of the allocated
buffers, ports coupled to a port on an other Local Manager take a large amount of memory, 1.1 MB.
Similarly, a native executable linked to
MUSCLE uses at least 650 kB of memory, and in Java an additional serialization
buffer is
allocated.
On a machine with 4 GB of memory per core, each core could accommodate up to 20,000
submodels with 10 local conduits each, up to 350 submodels with 10 remote
conduits, or up to 300 native submodels with 10 remote conduits. In most scenarios
this is more than sufficient, and the number of submodels will instead be limited by the
computational cost of the submodel code.

\begin{table}
\caption{Runtime and memory consumption\protect\footnotemark[2] of MUSCLE, on a local iMac and the PRACE
machine SuperMUC (see their details in Table~\ref{table:resources}). Entries marked `--' were not measured. We assume that memory
consumption on both machines is similar, since they both use 64-bit Intel processors.
The first row (Overhead)
indicates the overhead of MUSCLE without starting any submodels, the other rows
show additional overhead to this baseline.}\label{table:runtime}
\begin{tabular}{llll}
 \hline
& iMac runtime & SuperMUC runtime & iMac memory\\
 \hline
Overhead & 0.77 s & 1.2 s & 73 MB\\
Additional per submodel & 1.6 ms & 1.6 ms & 168 kB\\
Additional per local conduit & 0.11 ms & 0.10 ms & 3.4 kB\\
Additional per port with remote coupled port&--&--&1.1 MB\\
Additional per native submodel & 24 ms & -- & 1.7 MB\\
 \hline
\end{tabular}
\end{table}

\footnotetext[2]{Stated memory sizes are multiples of 1024 (kilo)bytes.}

\subsection{Message speed}\label{sec:messagespeed}

The performance of MUSCLE communication is compared with approaches that modellers would
usually use for composite models. The two most prevalent methods of communication
of our current users are file-based or MPI-based. The former is
often used to couple different codes together, whereas the latter is used to form
fast monolithic codes. For remote communication, GridFTP is a popular
alternative and MPWide is a well-optimized one. We will compare these methods
with the communication speed offered by MUSCLE~2. Both latency and throughput
of the methods will be computed.

\subsubsection{Single machine}\label{sec:messagespeed:local}

For the local communications we will compare speeds of file copy, MPI, MUSCLE 1
and MUSCLE~2. Each of the tests is done with message size 0 kB and $2^i$ kB, with $i$ ranging
from 0 to 16, which is up to 64 MB. Since MUSCLE 1 will not send messages larger
than 10 MB, its measurements are limited to $i$ ranging from 0 to 13 (8 MB). 
Per message size, a message is sent back and forth
100 times, so it makes 100 round trips. The time to send one message of a
certain size is calculated as the average over the round trip times, divided by
two. The latency is calculated as the minimum time to send a message.
The message times are then fitted to a linear curve $a x + b$ for message size $x$,
where throughput is calculated as $\frac{1}{a}$ and $b$ is taken as the latency.

For applications without a coupling library, a simple way to transfer data
from one process to another is to write to a file which another process may read.
The operating system might cache this file so that the read operation is fast.
This scenario is simulated by creating files as messages.
One round trip is taken as copying a file and copying it back using the systems
file copy, which is equivalent to writing and reading a file twice.

For a monolithic model, possibly with multiple substructures or threads, MPI is
a well-known and very fast option. This paradigm, however,
gives none of the plug and play advantages that MUSCLE~2 has, nor does it keep
time in sync between submodels, nor is it easy to combine resources of
different providers. In our experiment, messages are sent by one MPI process, then received and
sent back by another with the same executable.

To test MUSCLE, first we take the situation that all instances have a Java
implementation and a single machine is sufficient to run them. As described in
Section~\ref{sec:runtime}, messages are then sent through shared memory. Next, we take
two MUSCLE processes that communicate with TCP/IP, for
when a user wants to prioritize one process over the other, for instance. Finally,
we take two instances that both have a C++ implementation.

The file copy, MPI and MUSCLE~2 scenarios are tested on the iMac (local desktop),
Zeus (cluster), and SuperMUC (supercomputer). MUSCLE 1 is only
tested on the iMac due to portability issues.

\begin{figure}[ht]
  \centering
  \begin{subfigure}[b]{0.35\textwidth}
          \centering
          \includegraphics[width=\textwidth]{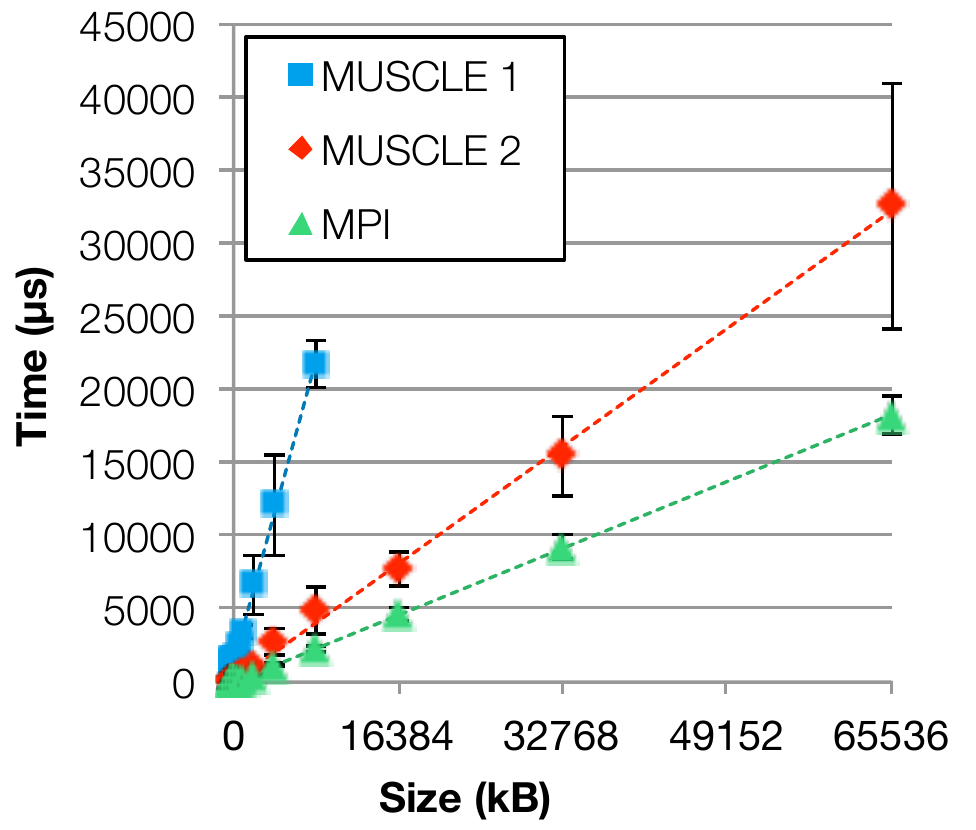}
          \caption{Local message time}
          \label{fig:local:performance}
  \end{subfigure}~
  \begin{subfigure}[b]{0.24\textwidth}
          \includegraphics[width=\textwidth]{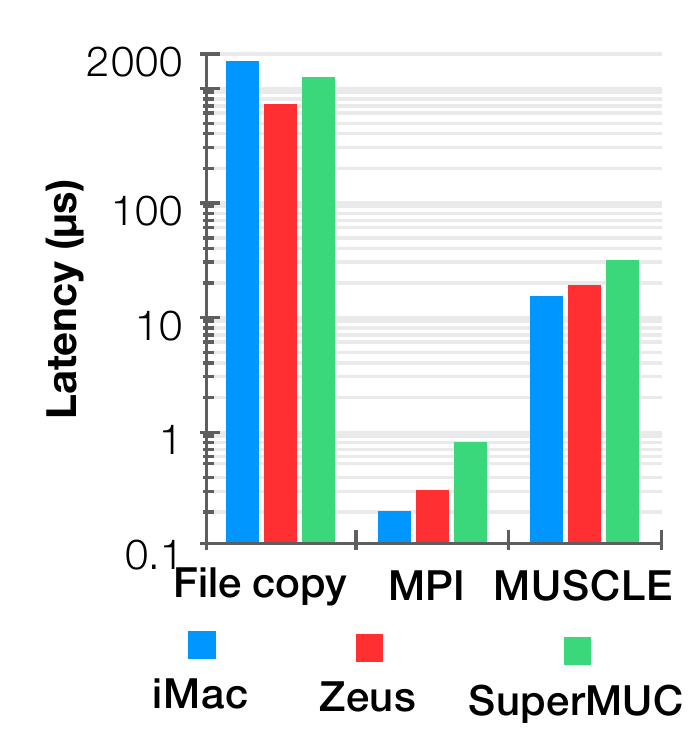}
          \caption{Local latency}
          \label{fig:local:latency}
  \end{subfigure}~
  \begin{subfigure}[b]{0.24\textwidth}
          \centering
		  \includegraphics[width=\textwidth]{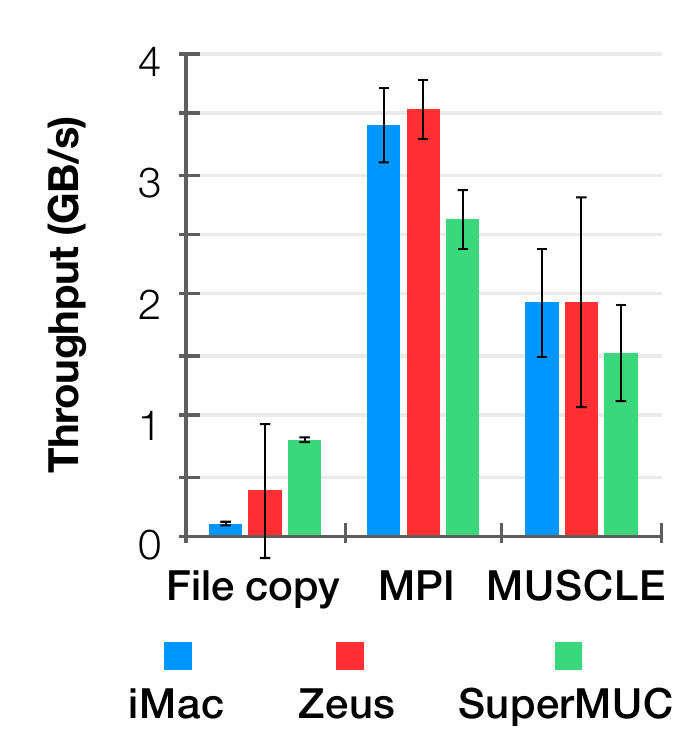}
          \caption{Local throughput}
          \label{fig:local:throughput}
  \end{subfigure}\\
  \begin{subfigure}[b]{0.24\textwidth}
          \includegraphics[width=\textwidth]{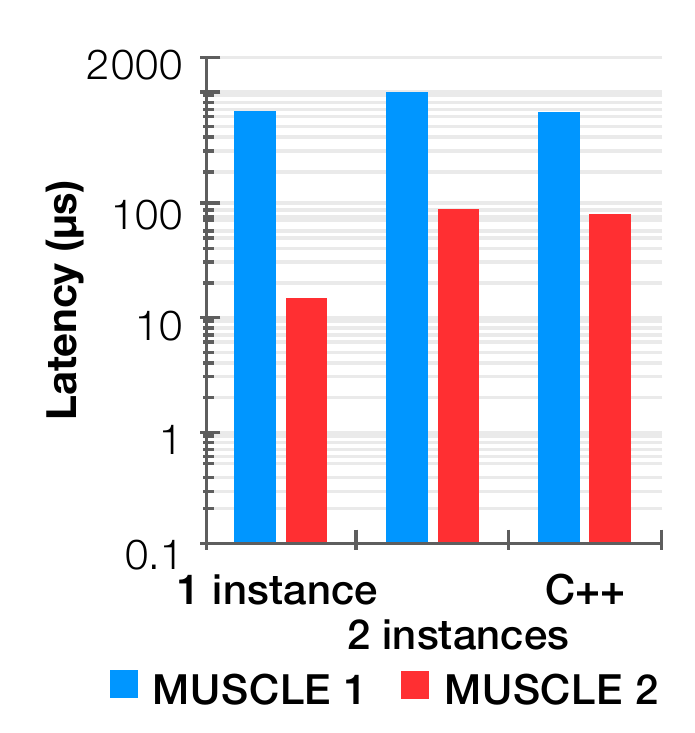}
          \caption{Run mode latency}
          \label{fig:muscle12:latency}
  \end{subfigure}~
  \begin{subfigure}[b]{0.24\textwidth}
          \centering
          \includegraphics[width=\textwidth]{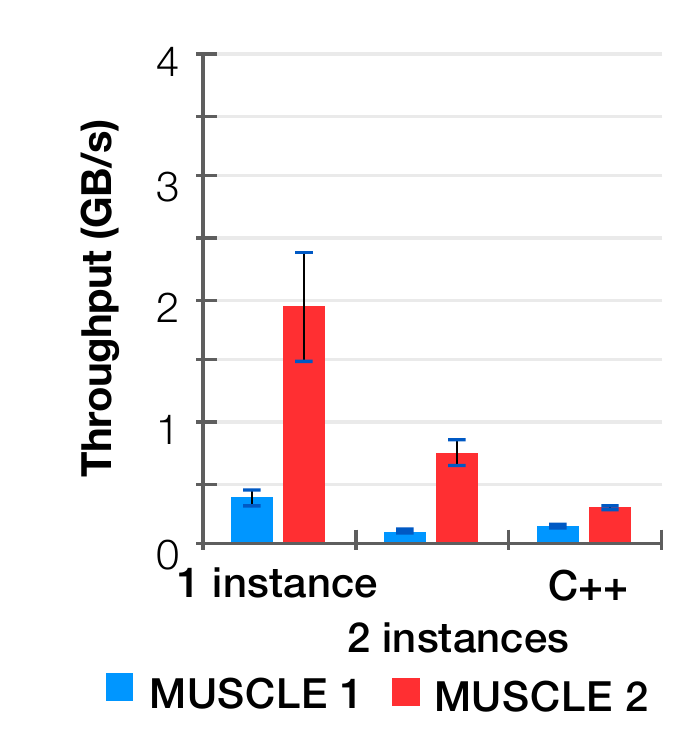}
          \caption{Run mode throughput}
          \label{fig:muscle12:throughput}
  \end{subfigure}
  

  \caption{The performance of the communication methods described in Section~\ref{sec:messagespeed}.
  (a) shows a plot of the time to send a single message, along with linear fit. 
  (b) and (c) show the performance of sending a message within a machine,
  for three machines (the latency is averaged as it showed little
  variation). (d) and (e) show the performance
  of sending messages on a local machine with MUSCLE~1 and MUSCLE~2,
  by starting a model: in a single MUSCLE instance; with two coupled MUSCLE
  instances; or, with C++ submodels. The standard error of the latency
  measurements was negligible so it is only shown for the throughput.}\label{fig:local}
\end{figure}

The results are plotted in Fig.~\ref{fig:local}. The standard deviation for
the latency is very low and is not shown. Obviously, copying data has a higher
latency and lower throughput than the alternatives. The latency
of MPI is clearly the lowest and MPI has the highest throughput as well, which
would be expected because it uses highly optimized native code. MUSCLE falls in
the mid-category, and is thus a serious contender if neither a monolithic nor
a file-based simulation is desired. These results do signal that for optimal
performance of a very tightly integrated code, MPI could be preferred over MUSCLE~2.
Of course, this MPI code can then be used in MUSCLE~2 as a single submodel, so that
MUSCLE can take care of starting the submodel and coupling it with other codes.

Comparing the run-modes of MUSCLE, and MUSCLE~1 and MUSCLE~2, a few
remarks can be made. First, the latency of C++ is lower than having two Local
Managers, which is surprising: with C++ a message is first serialized with XDR,
sent, passed through shared memory in Java and then serialized again to be sent
to another C++ program. With two Local Managers, a message is serialized once
with MessagePack and directly used. So although the throughput of MessagePack is
higher, its latency is worse. Second, MUSCLE~1 falls far behind MUSCLE~2 in all cases,
since it uses the JADE system to send messages and overall has a less optimized code.
That the performance of MUSCLE~1 is most similar to MUSCLE~2 in the C++ scenario,
is because the Java Native Interface (JNI) transfers data between Java and C++ faster
than TCP/IP sockets can. JNI was removed in MUSCLE~2 due to the
portability issues that it caused, see Appendix~\ref{appendix:muscle1} for the
details.

\subsubsection{Distributed computing}\label{sec:messagespeed:distributed}

Besides local message speed, distributed message speed is also important for
computing on large infrastructures. Although the main bottleneck will usually be the
available network bandwidth, software does have an influence on message
speed. In this section we will compare the speed of three possible technologies
to do wide area network communication: MPWide~1.8, GridFTP~0.8.9, and MUSCLE~2
with the MTO.
MPWide is designed specifically for optimally making use of the
available bandwidth by using packet pacing, multiple streams per connection and
adapted buffer sizes. GridFTP~\cite{GridFTP} is a dedicated file transfer
service run by EGI and PRACE sites. MUSCLE uses the MTO, which by default uses a
single plain TCP/IP socket per connected MTO but can also be used in conjunction
with MPWide.

Each test was performed between a PRACE Tier-1 site Cartesius in Amsterdam and
the PRACE Tier-0 site SuperMUC in Garching, Munich (more details in Table~\ref{table:resources}).
They send a message from
Amsterdam and back again, using message sizes 0 kB and $2^i$ kB, with $i$
ranging from 0 to 20, which is up to 1 GB. For each message size up to 1 MB, hundread
messages were sent, for messages ranging from 2 MB to 1 GB ten messages were sent.
The TCP/IP route from Cartesius to SuperMUC uses the high-speed PRACE network.
The average ping time over 50 consecutive pings on this route was 15.2 ms.


In all applications the standard settings were used. For MPWide the number
of streams must be specified and was set to 128 streams.
Although GridFTP can open multiple TCP streams for a transfer,
firewall settings prevented it to do so from SuperMUC to Cartesius, so in this
experiment it used only one.

\begin{figure}[ht]
	\centering
  \begin{subfigure}[b]{0.35\textwidth}
          \centering
          \includegraphics[width=\textwidth]{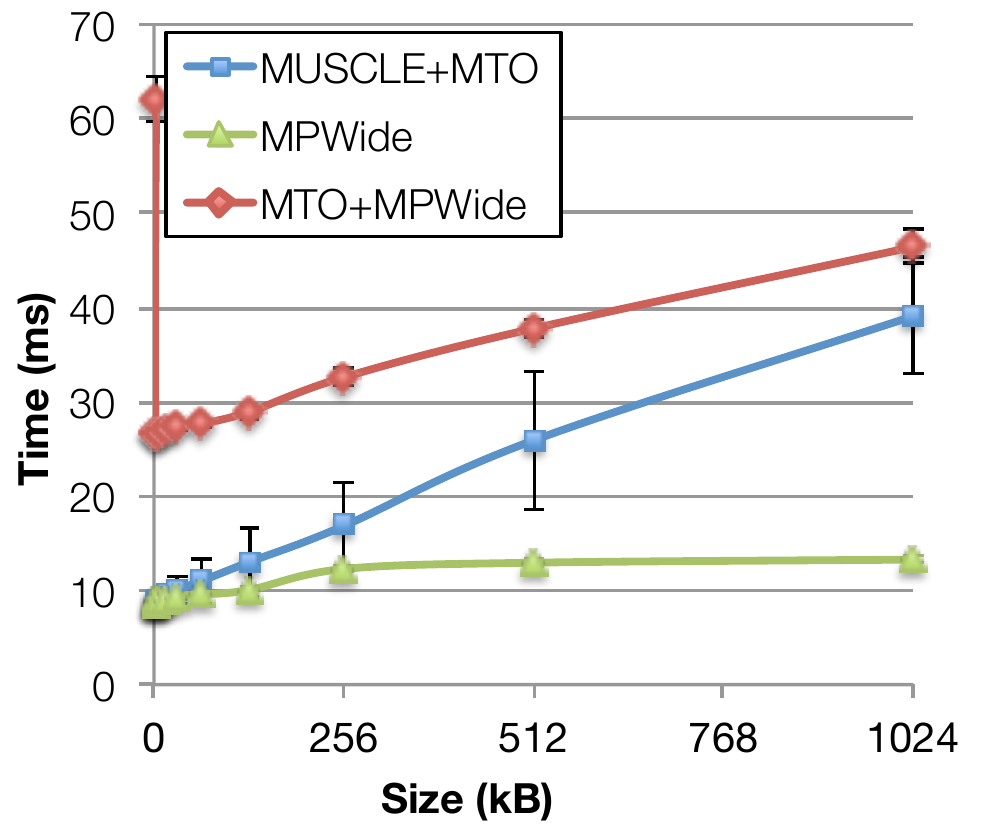}
          \caption{Distributed small message time}
          \label{fig:distributed:performance:kb}
  \end{subfigure}~
  \begin{subfigure}[b]{0.35\textwidth}
          \centering
          \includegraphics[width=\textwidth]{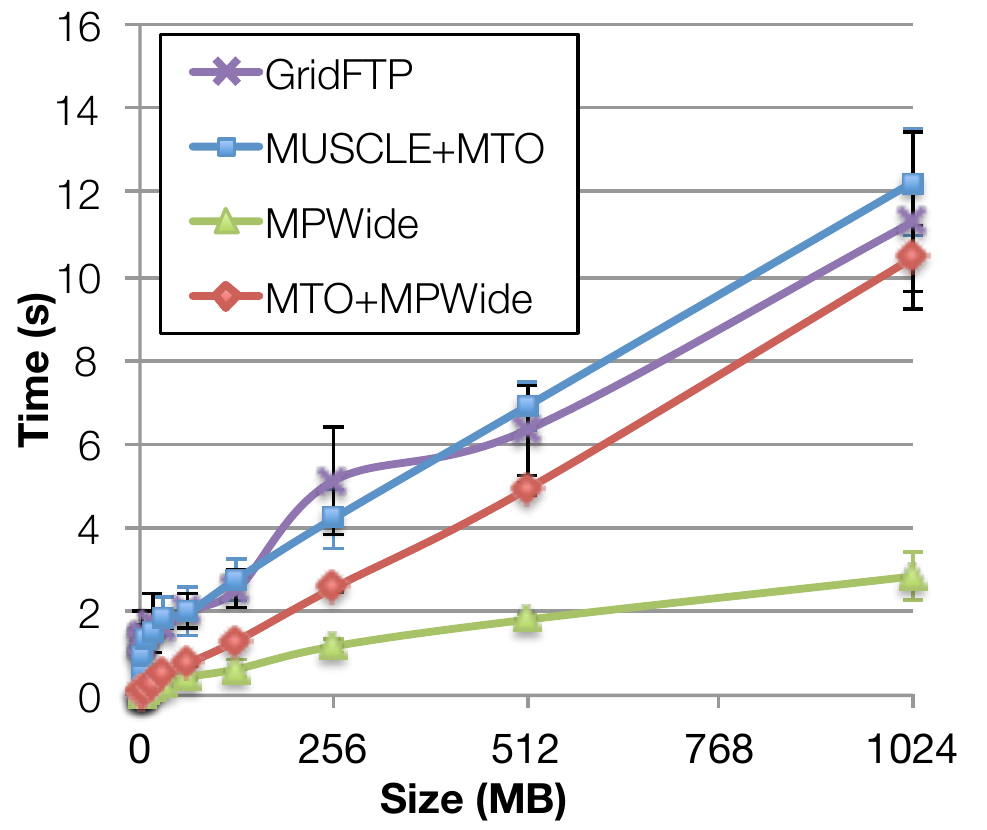}
          \caption{Distributed large message time}
          \label{fig:distributed:performance:mb}
  \end{subfigure}\\
  \begin{subfigure}[b]{0.29\textwidth}
          \centering
          \includegraphics[width=\textwidth]{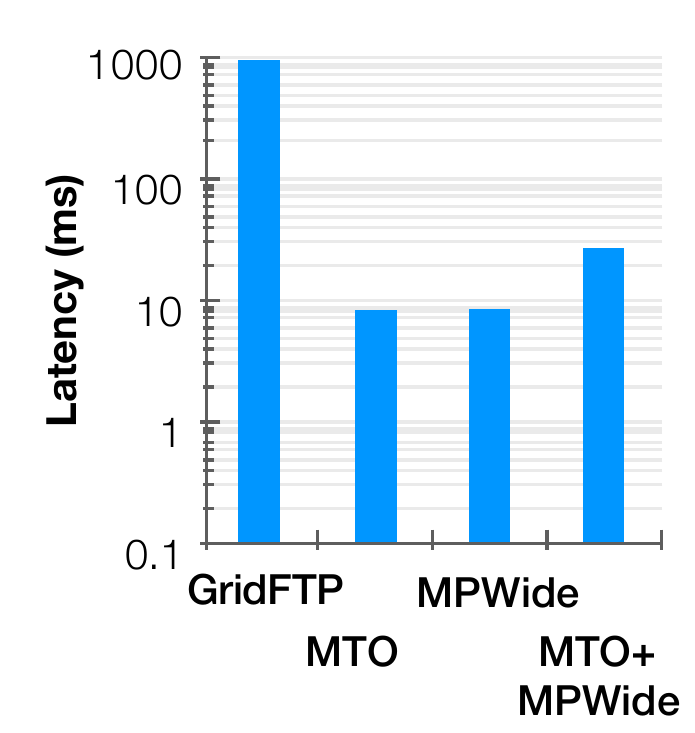}
          \caption{Latency}
          \label{fig:distributed:latency}
  \end{subfigure}~
  \begin{subfigure}[b]{0.29\textwidth}
          \centering
		  \includegraphics[width=\textwidth]{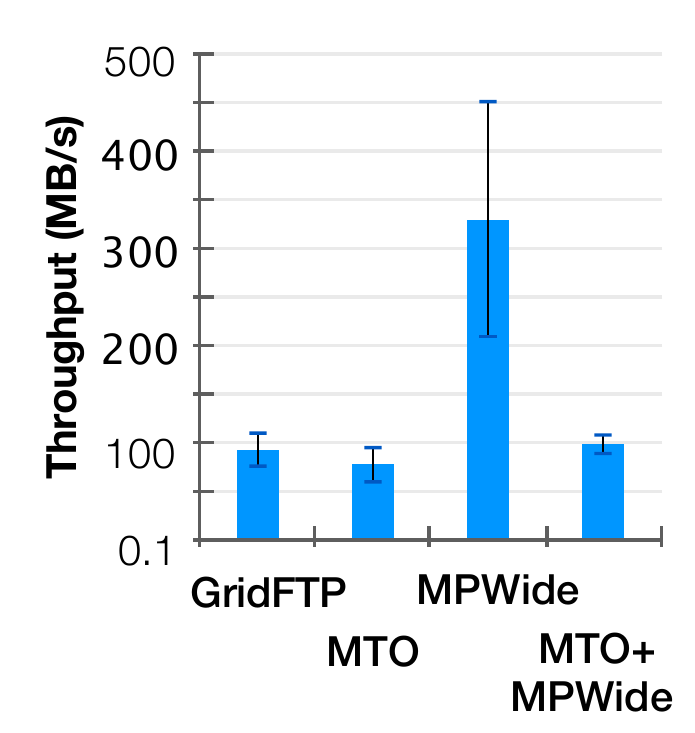}
		  \caption{Throughput}
          \label{fig:distributed:throughput}
  \end{subfigure}
  \caption{The performance of sending
  a message between Huygens and SuperMUC. (a) shows the time to send a message
  on a kilobyte scale, and excludes GridFTP which fluctuates around 890 ms.
  (b) shows the time to send a message on a megabyte scale.
  The other two plots show the fitted values of the data.}\label{fig:distributed}
\end{figure}

The results of the test are shown in Fig.~\ref{fig:distributed}. For very small
messages both MUSCLE~2 with the MTO and MPWide come very close to the ping time,
adding up to 2 ms. When the MTO uses MPWide internally the latency goes up
considerably because it uses an additional management layer. GridFTP has
to do a certificate hand-shake before when connecting, which takes significantly
longer at about 890 ms. With large messages its performance is much better, at 90
MB/s, although it does show an occasional bump when the hand-shake can not
be processed immediately. MUSCLE~2 with the MTO did a bit worse and MUSCLE~2 with
the MTO using
MPWide did a bit better. Plain MPWide performance was much better than the
other methods for messages larger than 128 kB. This indicates that further efforts
to integrate the MTO and MPWide may be beneficial.
%
\section{Use cases}\label{sec:usecases}

To show the real-time usage of MUSCLE~2 as well as its practical performance
we will show how it is applied to a multiscale model of a canal system and
specifically to the submodel of one canal section. Next, a multiscale model of
in-stent restenosis shows the heterogeneity of submodels that can be coupled.

\subsection{Hydrology application}\label{sec:canals}\label{sec:canals:performance}
\lstset{ 
  tabsize=1,
  numbers = left,
  breaklines=true,
  showstringspaces=true,
  basicstyle=\ttfamily,
  numberstyle=\scriptsize\sffamily,
  columns=fullflexible,
  keywordstyle=\color[rgb]{0,0,1},
  commentstyle=\color[rgb]{0.133,0.545,0.133},
  stringstyle=\color[rgb]{0.627,0.126,0.941},
  prebreak = \raisebox{0ex}[0ex][0ex]{\ensuremath{\hookleftarrow}},
  keywordstyle=\color[rgb]{0,0,1},
  framexleftmargin=10mm,
  xleftmargin=8mm,
  showstringspaces=false,
  }

\def\Lx{L_{x}}
\def\Lxk{L_{x,k}}
\def\Tserial{T_\text{serial}}
\def\Tmono{T_\text{mono}}
\def\Tmuscle{T_\text{muscle}}
\def\Tcom{T_{com}}
\def\dx{\Delta x}

\begin{figure}[tb]
\centering 
\includegraphics[width=0.8\textwidth]{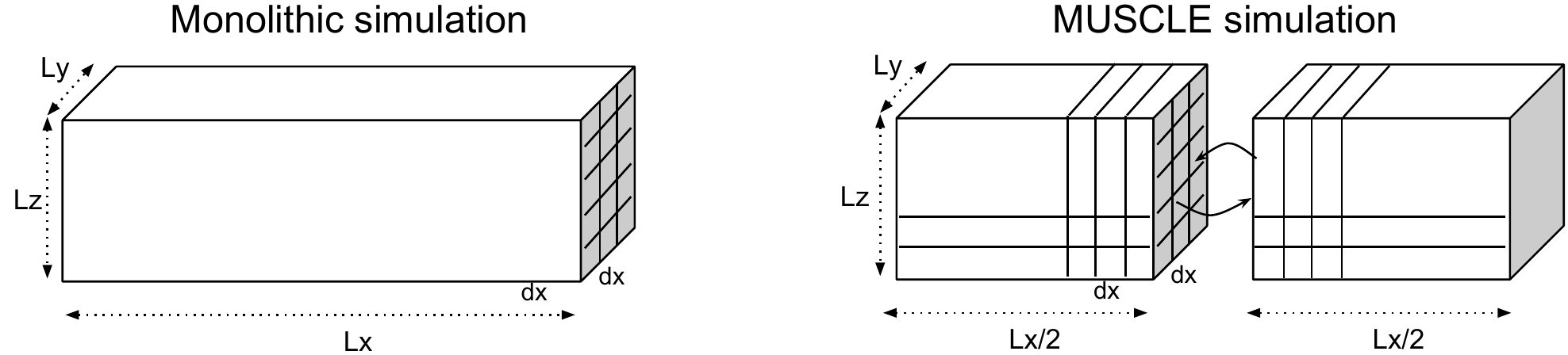}
\caption{How a 3D cavity can be split into equal parts for use with MUSCLE, where
$L_x$ is its length, $L_y$ its width and $L_z$ its depth, and $d x$ the resolution at
which it is resolved.}
\label{fig:split}
\end{figure}

An optimal management of rivers and waterways is a necessity
in modern society to ensure an adequate supply of water, in particular for
agriculture, electricity production or transportation~\cite{Marcou:2007kf}.
An important requirement is to control the water level and sediment transport in populated
areas~\cite{Malaterre:1998bn}. These problems can be addressed through computer simulation in
combination with optimization methods.

Many of such hydrology problems can be implemented using a ``Lego based
philosophy''~\cite{benbelgacem2013,BenBelgacem:2012iu}, where river or water
sections are modelled by submodels and connected with mappers, based on the
topology of existing canal systems. A submodel can for instance implement a 3D
Free Surface (3DFS) model and be connected to a 1D shallow water submodel. Because
of their different resolution and time step this gives a multiscale system. The
decomposition into submodels allows a distributed execution, which may be
necessary for larger canal systems.

Our use case consists of a 3D cavity flow problem solved with the Lattice
Boltzmann (LB)~\cite{succi:2001vh} numerical method. The submodels are implemeted with
the Palabos toolkit\footnote{Palabos: \url{http://www.palabos.org/}}, which uses MPI for
parallelisation. The aim is to evaluate the time overhead induced by the use of
the MUSCLE API when performing distributed computations of hydrodynamical
problems. As illustrated in Fig.~\ref{fig:split}, the computational
domain (here a 3D cavity) is divided across several parallel clusters and
information should be exchanged between them at each iteration. This use case
itself has full scale overlap but will be coupled to different time scales when
a canal system is simulated.

\begin{lstlisting}[float=htbp,language=c++,caption=Pseudo-code of the cavtiy3d example.,label=lst:code]
f_init()
for (iterations < maxIteration) {
    collideAndStream()
    gatherBoundaryData()
    sendReceiveBoundaryData()
    updateBoundaryData()
}
\end{lstlisting}

Listing~\ref{lst:code} gives the pseudo-code of the algorithm used by
the numerical method. 
During each loop iteration (line 2), the submodel computes the flow on line 3
using the parallel Lattice Boltzmann method. On Line 4, each submodel retrieves
boundary data from all MPI processes in the same job and submodel. Line 5
establishes the coupling between submodels. In this case, the
submodel sends and receives boundary data using the MUSCLE API hidden in the
\verb|sendReceiveBoundaryData()| function. On line 6, each section updates its
boundary according to the data received from the other submodels.

\def\u{\mathbf{u}}
\def\v{\mathbf{v}}
\def\x{\mathbf{x}}
\def\dt{\Delta t}
\def\dx{\Delta x}
\def\ov{\mathbf{\theta}}
\def\Tmono{T_\text{mono}}
\def\Tlocal{T_\text{local}}
\def\Tgrid{T_\text{distr}}
\def\egrid{\varepsilon_\text{distr}}
\def\elocal{\varepsilon_\text{local}}
\newcommand{\myparagraph}[1]{\paragraph{#1}\mbox{}\\}

To show the performance of MUSCLE when it is used in an actual 
problem, we will consider the performance of the 3D cavity submodel described
above. Our benchmark will consist of running a
monolithic code first and comparing its runtime with using two MUSCLE submodels.
A detailed treatment has been made by
Belgacem~\emph{et~al.}~\cite{benbelgacem2013}; here we show some results obtained
by using more CPUs.

The computational domain of the canal we will use, as depicted in Fig.~\ref{fig:split},
has a length $L_x$ of $13000$ metres, a width $L_y$ of $40$ metres and a depth $L_z$ of $10$ metres.
The spatial resolution $\dx$ may vary and will determine the problem size:
decreasing $\dx$ implies increasing the domain size.

For the benchmark, we will evaluate three scenarios:
\begin{enumerate}
\item a monolithic simulation of the canal on a single cluster;
\item a simulation with two canal submodels on the same cluster, coupled using MUSCLE; and
\item a simulation with two coupled canal submodels on different clusters, coupled using MUSCLE.
\end{enumerate}

The first case shows what the performance of a usual monolithic model with MPI
is, the second what the cost is in splitting that into multiple parts using MUSCLE,
and the third what the cost is of distributing it with MUSCLE.

The execution time of these scenarios is indicated $\Tmono$, $\Tlocal$, and $\Tgrid$, respectively.
In scenarios 2 and 3, the canal section computed in scenario 1 is split equally
amongst the submodels called left and right.
Each simulation carries out $100$ iterations, and this is repeated three times.
We varied the number of grid points per metre $N=\frac{1}{\dx}$ from 0.5 to 4,
with a step size of 0.5. For the total domain this means varying the problem size
from under 820 thousand grid points to over 340 million points, scaling with $N^3$.
The MUSCLE communication volume, however, only scales with $N^2$, so computation
will dominate computation for increasing $N$.

The simulations are run on the Gordias and Scylla clusters (for their details
see Table~\ref{table:resources}). The monolithic execution is done with 100
cores of the Gordias cluster. Likewise, the MUSCLE execution is done with 100
cores, but here the left
and right section run on 50 cores each. The local MUSCLE 
execution is run on Gordias whereas the in distributed one, both clusters are used.
In the local execution we first ran the
MUSCLE Simulation Manager in a separate node so that it had a fixed address before the job started.
In the distributed scenario, QCG-Broker takes care of queuing the jobs
and starting the Simulation Manager.
\begin{figure}
    \centering
    \begin{subfigure}{0.4\textwidth}
            \includegraphics[width=\textwidth]{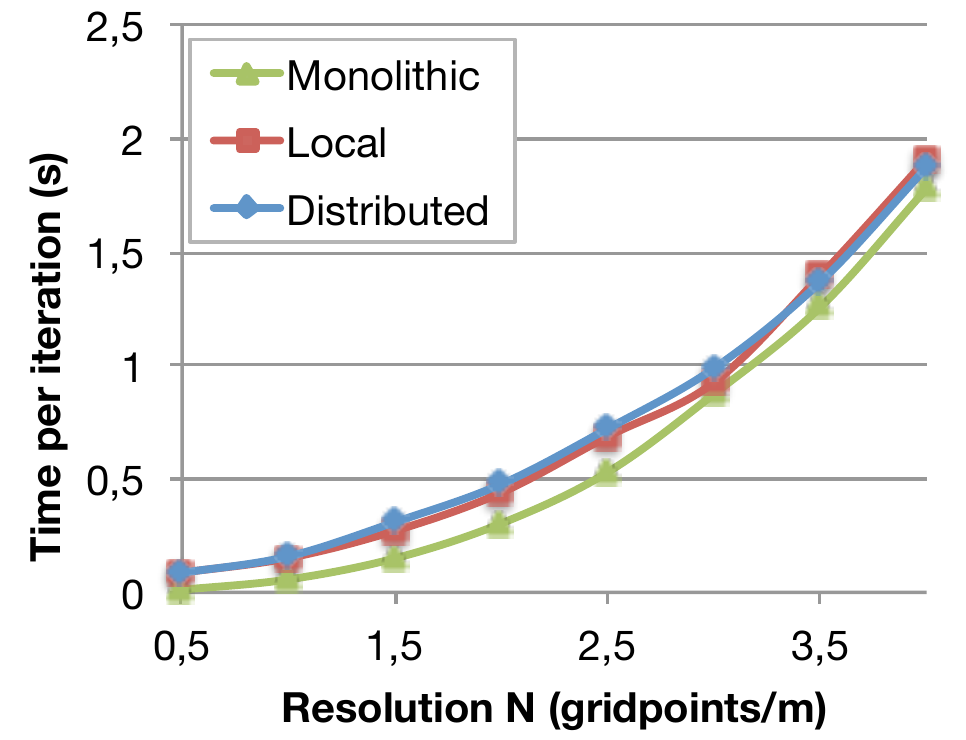}
            \caption{Runtime}
            \label{fig:canals:time}
    \end{subfigure}
    \begin{subfigure}{0.4\textwidth}
            \includegraphics[width=\textwidth]{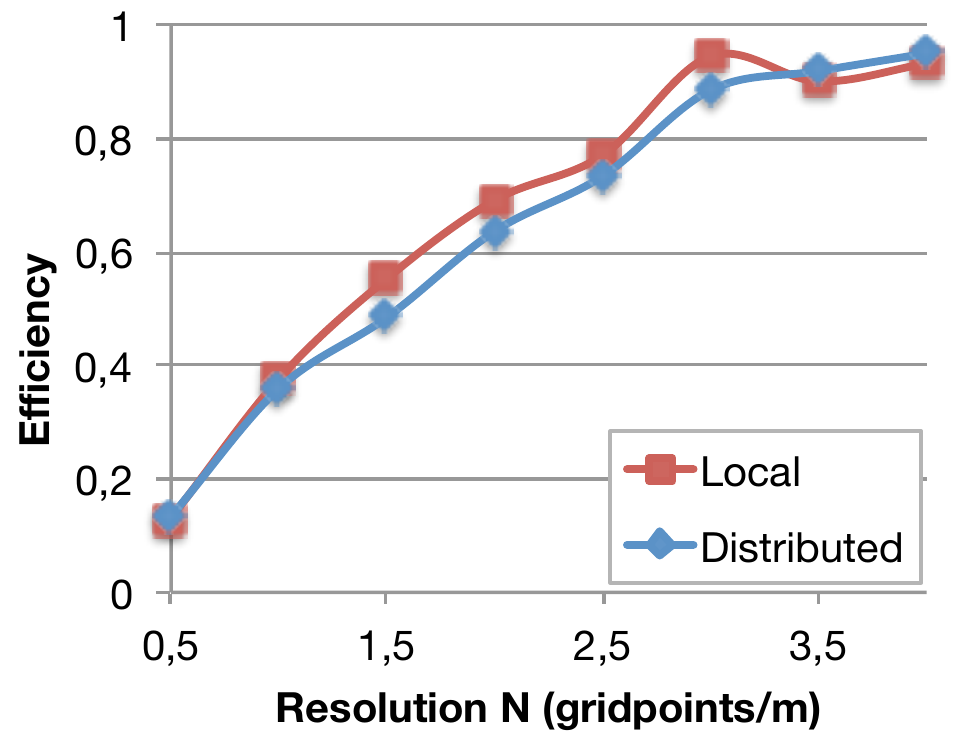}
            \caption{Efficiency compared to monolithic runtime}
            \label{fig:canals:efficiency}
    \end{subfigure}
    \caption{The performance of the three execution scenarios of the cavity 3D
    model, for the number of grid points per metre $N$, as described in
    Section~\ref{sec:canals:performance}.}
        
 \label{fig:canals:timeCompare}
\end{figure}

Fig.~\ref{fig:canals:time} shows the results of the benchmark of $\Tmono$, $\Tlocal$ and $\Tgrid$ 
on Gordias cluster. We measured the average time per iteration. If we compare
$\Tmono$ and $\Tlocal$, we see that the difference between execution times
varies very little over all values of $\dx$. The main bottleneck seems to be a
fixed synchronization overhead due to waiting for messages between submodels.

\begin{figure}[ht]
    \centering
    \begin{subfigure}{0.348\textwidth}
            \includegraphics[width=\textwidth]{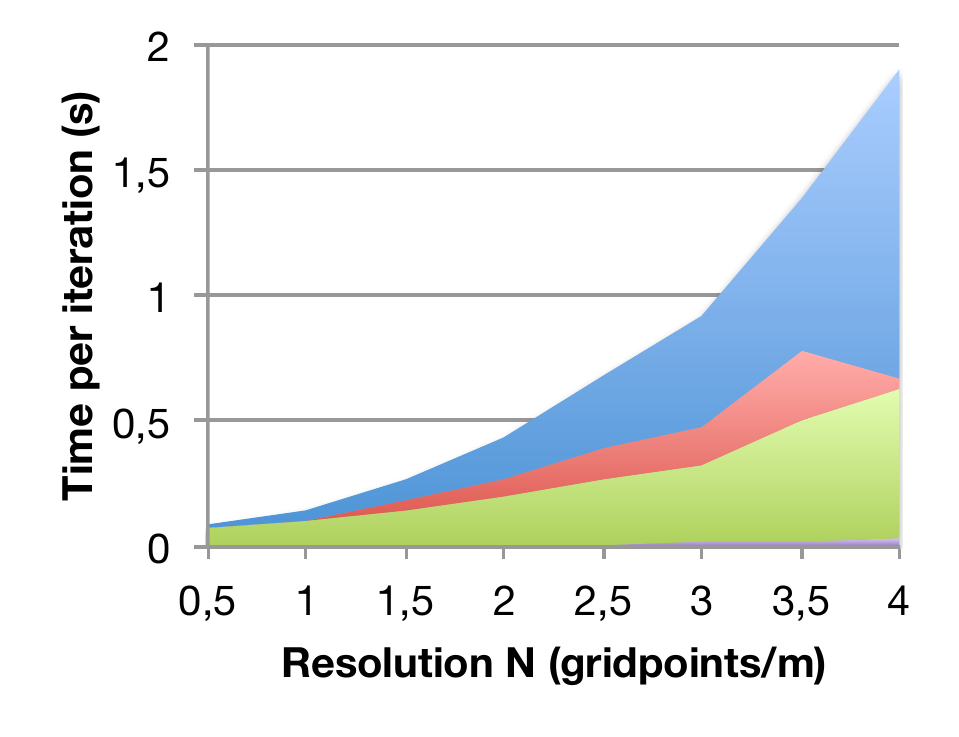}
            \caption{$\Tlocal$ of section 1}
            \label{fig:canals:intra_gordias1}
    \end{subfigure}
    \begin{subfigure}{0.452\textwidth}
            \includegraphics[width=\textwidth]{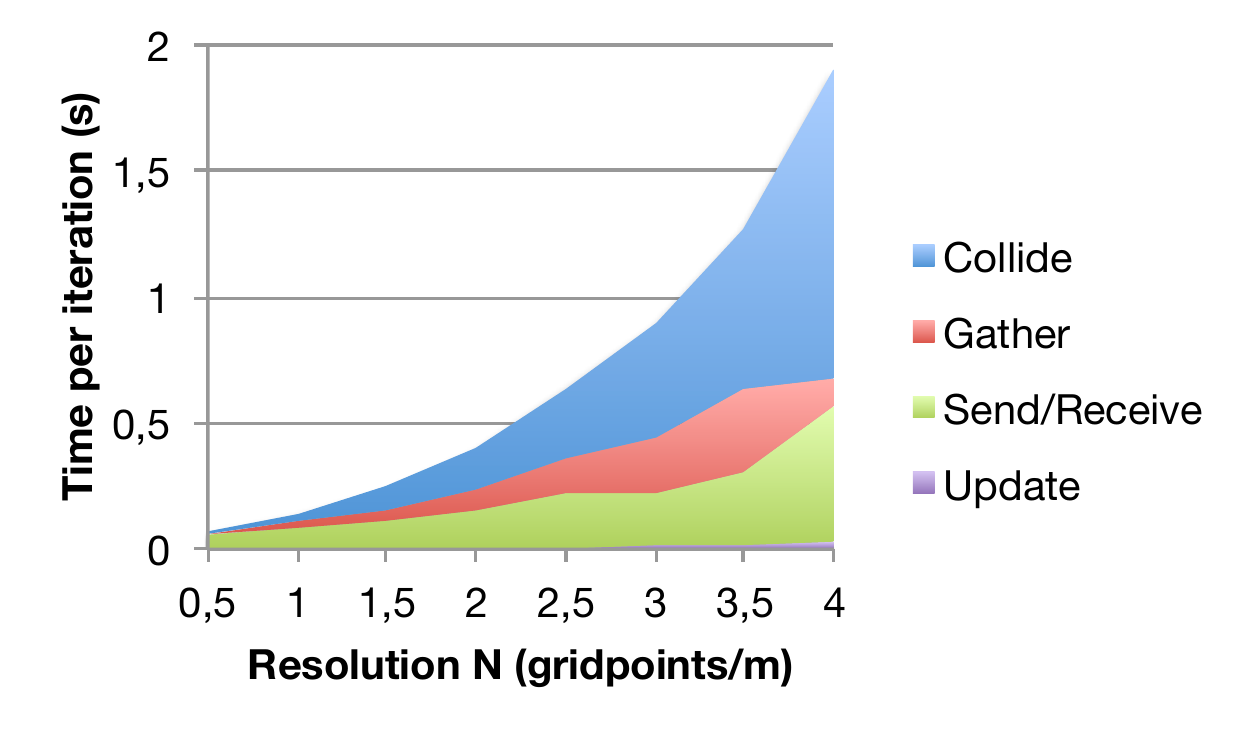}
            \caption{$\Tlocal$ of section 2}
            \label{fig:canals:intra_gordias2}
    \end{subfigure}
\\
    \begin{subfigure}{0.348\textwidth}
            \includegraphics[width=\textwidth]{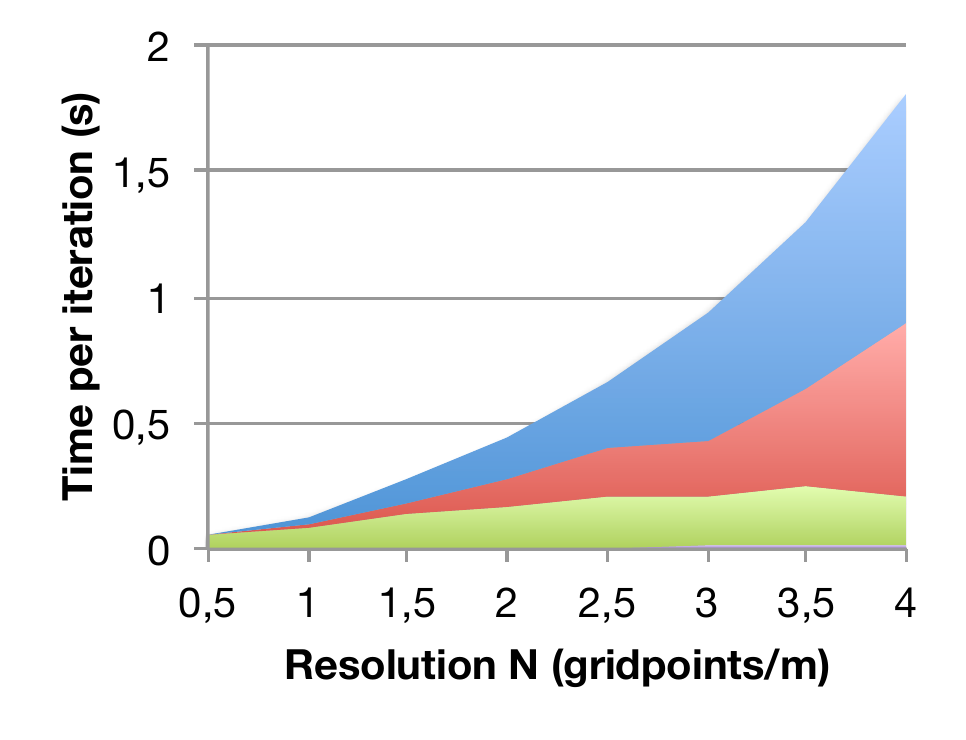}
            \caption{$\Tgrid$ of section 1}
            \label{fig:canals:inter_gordias}
    \end{subfigure}
    \begin{subfigure}{0.452\textwidth}
            \includegraphics[width=\textwidth]{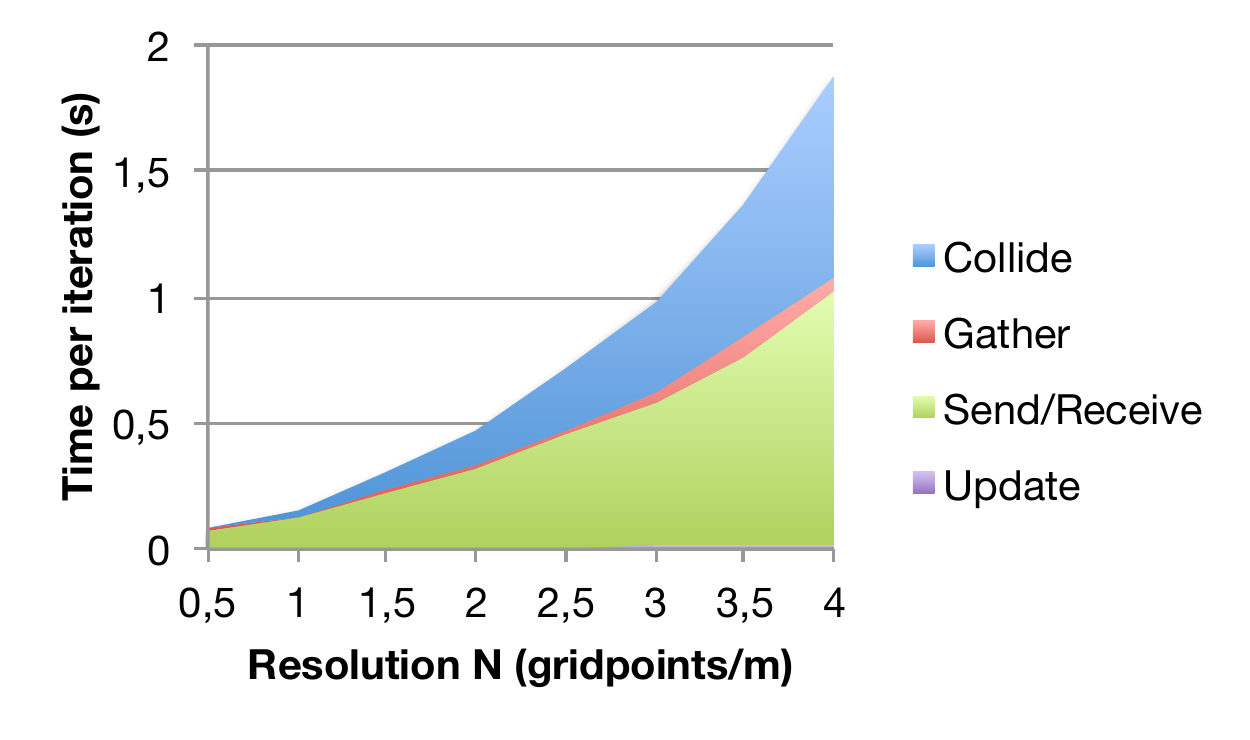}
            \caption{$\Tgrid$ of section 2}
            \label{fig:canals:inter_scylla}
    \end{subfigure}
    \caption{The runtime of different operations in the local and distributed cases, where
	(a)--(c) run on the Gordias cluster and (d) runs on the Scylla cluster, and (a)--(b) are run concurrently
	as are (c)--(d).
	The operations match the pseudo-code in Listing~\ref{lst:code}: \texttt{collideAndStream()} [Collide];
\texttt{getBoundaryData()} [Gather];
\texttt{sendReceiveBoundaryData()} [Send/Receive];
and \texttt{updateBoundaryData()} [Update].}
        
 \label{fig:canals:timeOperation}
\end{figure}

Regarding the distributed execution (Fig.~\ref{fig:canals:efficiency}), the efficiency values
$\egrid = \Tmono/\Tgrid$ and $\elocal=\Tmono/\Tlocal$ show the same behaviour;
i.e, we observe a large communication ratio with small values of $N$, and vice
versa. This goes to the extent that for $N=4$, using only MPI is just 5\%
more efficient than using MUSCLE.
$\egrid$ is smaller than $\elocal$ for smaller problem but for
large values of $N$ $\egrid$ is slightly higher, which can be explained
if we look at the detailed plots.

The runtimes of the two sections on the Gordias cluster, per operation of the pseudo-code~\ref{lst:code}, are very
similar, as shown in Fig.~\ref{fig:canals:intra_gordias1} and Fig.~\ref{fig:canals:intra_gordias2}.
For large $N$, the fraction of time spent actually calculating increases steadily.
For smaller $N$, however, most of the time is spent in waiting for messages from
the other submodel, so if one submodel was slower then the other would have to wait until it was
finished and vice-versa. In the distributed experiment however, the submodel on Scylla (Fig.~\ref{fig:canals:inter_scylla})
was computed consistently faster, which means the submodel on Gordias
(Fig.~\ref{fig:canals:inter_gordias}) needs to wait far less. This gives a lower
average time per iteration for situations that depend more on computational time
than on communication time.

\subsection{ISR3D}

The three-dimensional model of in-stent restenosis in coronary arteries (ISR3D),
covered in
\cite{Borgdorff:2012ge,Groen:2013jn} and first described in
\cite{Caiazzo:2011jl}, originally used MUSCLE 1 which was replaced by MUSCLE~2.
In-stent restenosis is the recurrence
of stenosis after stenting a stenosed blood vessel. The model is multiscale in its time
scale, where smooth muscle cells
proliferation in the blood vessel wall is modelled on a time scale of hours to
months, and the blood flow and shear stress is computed on a time scale of
milliseconds to a second. It couples a submodel using the C++ API with OpenMP
to another using C++ with MPI, and uses Fortran and Java in other submodels and mappers.
ISR3D has routinely been executed between sites in different countries using
MUSCLE with the MTO, running the parallel submodels on a highly parallel
machine and the serial parts on another site.
When the original custom blood flow submodel code needed to be replaced with the Palabos
library, the plug-and-play character of MUSCLE~2 proved useful, since other submodels
did not have to be altered in this operation.
\section{Conclusions}

In this contribution, we have introduced and discussed the component-based and
flexible design of MUSCLE~2, and its distributed computing capabilities.
It is based on a general approach to multiscale modelling and simulation~\cite{Borgdorff:2012uq,Hoekstra:2007er,Hoekstra:2010uv}
combined with the multiscale modelling language~\cite{Falcone:2010to,Borgdorff:2012uq}. Because of its
modular setup, clearly
separating API, coupling, and runtime environment, users can
modify parts of a multiscale model without affecting the rest. A multiscale
model implemented with MUSCLE~2 can be executed on distributed
computing resources at any stage. Moreover, submodel code written in
Java, C, C++, Python, or Fortran, and using serial code, MPI, OpenMP, or threads
can freely communicate with other submodels using different technologies.

The overhead of starting MUSCLE~2 for multiscale models with a reasonable amount
of submodels is shown to be low, both time- and memory-wise. For local computing
MUSCLE~2 is shown to be more efficient than file based message passing, but it
has a factor two lower throughput than MPI and up to 30 $\mu$s higher latency.
For parts of a multiscale model where MPI is better suited, such as performing
a lattice method or doing agent based simulations, MUSCLE~2 can simply run that
part as a submodel with MPI, and the multiscale model will still have the
advantages of flexible coupling and execution.

For distributed computing, the MUSCLE Transport Overlay transfers
data from one high-performance computing centers to another. Its efficient
transfers easily surpass GridFTPs speed for smaller messages and give performance
similar to GridFTP for large messages. Using MTO with MPWide gives slightly better
performance on the high-speed PRACE network, but plain MPWide still much faster, so the
integration between the MTO and MPWide will be further examined.

For a canal system model, MUSCLE~2 makes it easier to generate canal
topologies by flexible coupling and being able to distribute different parts of
the canal system. Moreover, for canal sections with sufficiently large problem sizes, 
the performance of MUSCLE~2 is competitive with using a single
monolithic code. It will need distributed computing for larger problems
when a local cluster does not provide enough resources; this turns out not to be
very detrimental to performance.




\section*{Acknowledgements}
We would like to thank Jules Wolfrat and Axel Berg at SurfSARA, Amsterdam for
providing access to the Huygens and Cartesius machines. The work made use of
computational resources provided by PL-Grid (Zeus cluster) and by hepia in Geneva
(Gordias cluster).

This research presented in this contribution is partially supported by the
MAPPER project, which receives funding from the EU's Seventh Framework
Programme (FP7/2007-2013) under grant agreement N$\ensuremath{^\circ}$
RI-261507.

\bibliographystyle{model1b-num-names}
\bibliography{library,library_misc,extra}

\appendix
\renewcommand*{\thesection}{\Alph{section}}

\section{Technical details of the MUSCLE~2 runtime environment}\label{appendix:technical}

To increase the separation between the model and the runtime environment each
mapper or submodel has its own instance controller that will do the actual communication
with other parts of the simulation. When an instance controller starts up it
first tries to register to the Simulation Manager. It then queries the Local
Manager for the location of all the instances that it has a sending conduit to.
The Local Manager will then query the Simulation Manager in a separate thread
if it does not know the location. When an instance is finished, its instance
controller will deregister it at the Simulation Manager.

Although each instance controller and thus each instance uses a separate thread
by default, it is also possible to implement submodels asynchronously. MUSCLE~2
will be able to manage a large number of light asynchronous submodels in a small
number of threads. This leads to both lower memory usage and faster computation
since there are far fewer thread context switches but it makes the submodel code
slightly more complex and, if not properly coded, prone to race conditions.

Error handling, throughout the program, is designed to work fail-fast. If an
uncaught exception occurs in one instance, MUSCLE~2 assumes that continuing the
simulation will not yield valid results and it will try to shut down all other
instances. This behaviour was implemented to prevent wasting resources
on systems that charges end users for the total wall-clock time used by a
simulation. It also prevents deadlocks when an instance still expects data from another that has
already quit. MUSCLE 2 does not provide error recovery, instead each submodel should handle its own
checkpointing, if needed.


\subsection{Implementation of the MUSCLE Transport Overlay (MTO)}

The MUSCLE~2 Transport Overlay (MTO) is a C++ user space daemon. It listens for
connections from MUSCLE 2 on a single cluster, and keeps in contact with MTO's
on other clusters. It forwards any data from MUSCLE 2 intended for another
cluster to that clusters MTO. To identify the MTO associated to a MUSCLE~2
TCP/IP address, each MTO mandates a separate port range to MUSCLE~2.

The default connection between MTO's uses plain non-blocking TCP/IP sockets, and
this is well tested. To optimize speed over wide area networks, it has a local buffer
of 3 MB and it will prefer sending over receiving up to the point that it will
not allow more incoming data if the send buffers are too large or numerous.
The MPWide~1.8~\cite{Groen:2010gw} library is optionally enabled for
connections between MTO's. MPWide is a library to optimize message-passing
performance over wide-area networks, especially for larger messages. This option
currently only works between a pair of MTO's and the performance depends on the
connection between the clusters, but there are ongoing efforts to increase the
compatibility.

\subsection{QosCosGrid and MUSCLE~2 integration}

We identified two main integration points of the QosCosGrid software stack and
MUSCLE~2. First, the location (IP address and port) of the MUSCLE Simulation
Manager can be exchanged automatically with other MUSCLE Local Managers via the
QCG-Coordinator service - a global registry that offers blocking call
semantics. Moreover, this relaxes the requirement that the Simulation Manager
and Local Managers must be started in some particular order. The second benefit
of using the QosCosGrid stack with MUSCLE is that it automates the process of
submission of cross-cluster simulations by: co-allocating resources and
submitting on multiple sites (if available, using the Advance Reservation
mechanism); staging in- and output files to and from every system involved in a
simulation; and finally, allowing users to peek at the output of every submodel
from a single location.

\subsection{Comparison between MUSCLE 1 and MUSCLE 2}\label{appendix:muscle1}

The largest changes in MUSCLE since MUSCLE 1 involve decoupling
functionalities. The separation between the library and runtime environment
makes the system more usable, since users now do not need to go through
MUSCLE internals to do basic operations like getting model parameters, and
this in turn makes submodel code less susceptible to being incompatible with
newer versions of MUSCLE. The separation of C/C++/Fortran code from the main
Java code makes compilation much more portable. Finally, the separation of
message passing code and the communication method allows choosing more
efficient serialization and communication methods when able.

In terms of portability, MUSCLE~2 comes with all Java prerequisites so they do
not have to be installed manually. Moreover, the number of required Java libraries
has been drastically reduced. Notably, MUSCLE~2 no longer relies on the Java Agent
Development Environment (JADE) for its communication. This way, the MUSCLE~2
initialization sequence and communication routines are more
transparent, which in turn lead to numerous performance enhancements to
communication protocols and serialization algorithms.
As a result, MUSCLE~2 can handle messages up to a gigabyte, while MUSCLE 1 will
not handle messages larger than 10 MB.
Although distributed execution was already possible with MUSCLE 1, it only
worked for specifically set up environments, whereas MUSCLE 2 will run with most
standard environments.

In MUSCLE~1 the Java Native Interface (JNI) was used to couple native
instances. Although JNI is an efficient way to transfer data from and to Java,
it gave MUSCLE~1 usability and portability issues and introduced
incompatibilities with OpenMP and MPI. In MUSCLE~2, submodels must link to the
MUSCLE~2 library instead, at a penalty of doing communications between Java and
C++ with the somewhat slower TCP/IP.

Additional new features of MUSCLE~2 include a CMake-based build system, having
standardized and archived I/O handling, more flexible coupling, and automated
regression tests.

\end{document}